# Evaluating human-AI workflows for field research in viticulture

Niko Carvajal Janke[a,*], Daoyuan Jin[b], Shivranjani Baruah[a], Nicholas Gunner[c], Jacob Maus[d], Yu Jiang[c], Kaitlin M. Gold[a]

[a] Plant Pathology and Plant-Microbe Biology Section, School of Integrative Plant Science, Cornell AgriTech, Cornell University, Geneva, New York, 14456, USA

[b] School of Electrical and Computer Engineering, Cornell University, Ithaca, New York, 14850, USA

[c] Horticulture Section, School of Integrative Plant Science, Cornell AgriTech, Cornell University, Geneva, New York, 14456, USA

[d] Colinas Farming Company, Rutherford, California, 94573, USA

*Corresponding author. Niko Carvajal Janke, nc572@cornell.edu

## Abstract

We assessed the value of two live human-AI interactions in a precision disease control project in California vineyards. The project tested whether 2021-2024 commercial scouting records and remote-sensing measurements across 140 hectares could support 2025 red-leaf symptom forecasting for prioritized scouting and virus testing. In Workflow 1, Aleks v1, a multi-agent research system, developed forecasting models with iterative human refinement. We applied Aleks's 2024 vine-scale model to updated 2025 predictors and evaluated red-leaf forecasts against independent 2025 scouting. In retrospective simulations surveying 45% of all vine positions, adding model-informed row prioritization to adaptive scouting increased the encountered proportion of newly recorded red-leaf observations from 85.8% to 94.1%. Within-block scouting comparisons suggested the model mainly improved scouting allocation among blocks. Despite unreliable internal 2024 performance estimates from synthetic oversampling before train/test splitting, Aleks developed an informative vine-scale model in 145 minutes, increasing throughput and answering our research questions. In Workflow 2, we assessed whether higher model-score vines had more frequent virus detection, and whether Aleks could infer this sampling goal from a general prompt with data and literature. Aleks's plan prioritized balanced vineyard and model score coverage, while our plan prioritized field efficiency and high-model-score oversampling. Aleks's and our plans yielded 41/50 (82%) and 97/100 (97%) sampled vines. Aleks's plan omitted instructions for replacing missing vines, limiting implementation and operational value. Five of 137 sampled vines tested positive for grapevine red blotch virus (model score ROC AUC 0.735). These findings support assessing AI interactions by how well they advance field research objectives under live, project-specific constraints. Further evaluations could refine protocols for reliable support across field research projects.

# 1 – Introduction

## 1.1 – Artificial intelligence systems for agricultural field research

Artificial intelligence (AI) research systems are defined here as software designed to independently complete research tasks, including literature synthesis, experimental design, and data engineering. Laboratory applications of AI research systems have enabled drug optimization and materials discovery at throughputs that would be difficult for human teams alone (Gridach et al., 2025; Pyzer-Knapp et al., 2022; Tobias & Wahab, 2025). However, the relevance of these deliverables depends on whether project-specific context is effectively communicated to and incorporated by the AI (Gridach et al., 2025). In agricultural field research, narrow sampling windows and site-specific variability can compound time and resource constraints during project design and execution (Araus et al., 2018; DeBruin et al., 2025; Paccioretti et al., 2025). AI systems have shown potential to address such bottlenecks by completing data-intensive tasks semi-independently, such as plant-phenotyping analyses through natural-language instructions (F. Chen et al., 2026) and completing selected agricultural data-management tasks faster than human users (Pan et al., 2024). To demonstrate that AI systems can address these bottlenecks during live field research, human-AI interactions need to be evaluated based on how they contribute to subsequent activities within an ongoing field research project.

Human-AI workflows built around recurring sequences of research tasks can provide a consistent structure for evaluating AI support across projects, AI systems, and forms of human interaction. To evaluate the value of AI involvement in a research workflow, research teams can specify the objective of each AI task, the constraint it addresses, the context and inputs introduced, and the criteria for an operationally meaningful deliverable (Gonzalez et al., 2026; Lekadir et al., 2025). This has been demonstrated through live case studies in chemistry and biomedical research, where AI-generated deliverables have been evaluated from creation through experimental implementation, capturing operational relevance beyond that represented in simulations or standalone task assessments (Boiko et al., 2023; Ghareeb et al., 2026). Applied to agricultural field projects, live workflow evaluations could test whether AI task delegation expands research capacity and whether resulting outputs remain useful under varying seasonal, site-specific, and commercially relevant project constraints.

The potential value of current AI research systems varies across agricultural research disciplines, with data-intensive disciplines offering substantial opportunities for AI support (Gridach et al., 2025; Kuska et al., 2024; Williamson et al., 2023). Research on disease in commercial specialty crops is one such discipline (National Academies of Sciences, Engineering, and Medicine, 2025; Portela et al., 2024; Tardaguila et al., 2021). In precision viticulture, sensing platforms and digital management systems increasingly enable site-specific disease detection and prediction using spatial, temporal, biological, and management data (Bono et al., 2026; Mikaberidze et al., 2025), but integrating their outputs can create data-engineering bottlenecks that narrow the questions research teams and vineyard managers can address (Lacoste et al., 2022, 2025; Song et al., 2022; Williamson et al., 2023). These bottlenecks are compounded by narrow observation and sampling windows shaped by seasonal symptom expression and commercial vineyard operations (Cieniewicz & Fuchs, 2025; Hobbs et al., 2023). Human-AI workflows may therefore help research teams in precision viticulture evaluate a wider range of modeling, validation, and sampling designs to address research and management bottlenecks.

### 1.2 – Mitigating grapevine viruses in California viticulture

Grapevine red blotch disease (GRBD) and grapevine leafroll disease (GLD) are economically important viral diseases of grapevine in California (Cieniewicz & Fuchs, 2025; National Academies of Sciences, Engineering, and Medicine, 2025). Modeled losses in California vineyards can reach $68,548 per hectare for GRBD and $226,405 per hectare for GLD over a 25-year vineyard lifespan (Ricketts et al., 2015, 2017). GRBD is caused by grapevine red blotch virus (GRBV), whereas GLD is associated with several grapevine leafroll-associated viruses, of which grapevine leafroll-associated virus 3 (GLRaV-3) is the most prevalent (Cieniewicz & Fuchs, 2025; Fust et al., 2025; National Academies of Sciences, Engineering, and Medicine, 2025). GRBV and GLRaV-3 can enter vineyards through infected propagation material and spread between vines through insect vectors (Cieniewicz & Fuchs, 2025; Flasco, Hoyle, et al., 2023; Fust et al., 2025). With no curative treatments, management includes virus-free planting material and removal or replacement of infected vines (Cieniewicz & Fuchs, 2025; Fust et al., 2025; National Academies of Sciences, Engineering, and Medicine, 2025).

Foliar symptoms vary with phenology, cultivar, and time since infection, and may overlap between GRBD and GLD and other biotic and abiotic conditions (Cieniewicz & Fuchs, 2025; DeShields & Kc, 2023; Fust et al., 2025; National Academies of Sciences, Engineering, and Medicine, 2025; Rohrs et al., 2025). Given this symptom variability and overlap, visual scouting alone cannot always establish whether a red-leaf observation is associated with GRBV, GLRaV-3, or another cause (Cieniewicz &

Fuchs, 2025; National Academies of Sciences, Engineering, and Medicine, 2025; Rohrs et al., 2025). GRBV and GLRaV-3 can also be present without visible symptoms, allowing infected vines to remain undetected sources of inoculum for vector-mediated spread (Flasco, Cieniewicz, et al., 2023; Flasco et al., 2025; Flasco, Hoyle, et al., 2023; Fust et al., 2025). Determining whether red-leaf symptoms are associated with GRBV or GLRaV-3 therefore relies on virus-specific testing (Cieniewicz & Fuchs, 2025; National Academies of Sciences, Engineering, and Medicine, 2025). Although laboratory testing can detect both viruses, the cost and labor required for testing can limit comprehensive vineyard surveys, and the timing and methods of testing can impact the reliability of negative results (Hobbs et al., 2023; National Academies of Sciences, Engineering, and Medicine, 2025; Rohrs et al., 2025). Given the cost and uncertainties involved in mitigating GRBD and GLD in commercial viticulture, there remains a need for decision support approaches to help prioritize finite scouting and molecular testing resources (Cieniewicz & Fuchs, 2025; Hobbs et al., 2023).

### 1.3 – Longitudinal vineyard records for site-specific decision support

Across perennial specialty crops, spatially explicit management records can track environmental conditions, management actions, and disease observations for individual plants or field blocks (Bührdel et al., 2020; Dhonju et al., 2024). In viticulture, mapping platforms can combine these records with aerial imagery and georeferenced field observations to support site-specific vineyard management decisions (Dhonju et al., 2024; Tardaguila et al., 2021). Spatially explicit symptom and test records can inform GRBD and GLD management strategies (Flasco et al., 2025; Rohrs et al., 2025), but scouting labor and virus-testing capacity can limit their coverage across a vineyard or property (Bell et al., 2021; Hobbs et al., 2023; National Academies of Sciences, Engineering, and Medicine, 2025). For GRBV, zonal roguing uses infection distributions to define removal zones around vines likely to harbor infections (Flasco et al., 2025), whereas adaptive cluster sampling expands testing around positive vines to resolve local infection clusters (Rohrs et al., 2025), making both approaches dependent on spatially resolved infection data. GLD models likewise show that local disease pressure and production economics drive preferred combinations of vine removal, replacement, and vector control for growers, underscoring the need to match virus management to property-specific conditions and priorities (Bell et al., 2021; Ricketts et al., 2015).

Models combining grower records and remote sensing could help prioritize symptom scouting and virus testing across a vineyard management area (Colaço et al., 2025; Dhonju et al., 2024; Mikaberidze et al., 2025; Tardaguila et al., 2021). Remote sensing can support vine-scale disease detection, but performance can vary across cultivars, vineyards, and growing conditions, creating a need for vineyard-specific calibration and evaluation (Eeckhout et al., 2026; Laroche-Pinel, Cooper, et al., 2025; Laroche-Pinel, Singh, et al., 2025; Mikaberidze et al., 2025). Longitudinal grower records can provide site-specific management and disease histories that can contextualize remotely sensed variation in vine condition and observed reflectance patterns, creating an opportunity to integrate these complementary data into site-specific decision-support models (Bührdel et al., 2020; Colaço et al., 2025; Mikaberidze et al., 2025). In this case study, we asked whether longitudinal vineyard management records and remote-sensing products could support site-specific red-leaf symptom forecasting for prioritized red-leaf scouting, and whether higher model scores were associated with more frequent virus detection. We organized the research tasks needed to answer these questions into two human-AI workflows to assess the potential value of AI-supported model development and project-specific sampling plan design.

## 2 – Methods

### 2.1 – Evaluating AI support through live research workflows

We evaluated Aleks v1, an AI research system (Jin et al., 2025), in two live field research workflows (Figure 1). Here, a workflow was a coordinated sequence of human and AI tasks used to answer a research question. We evaluated our interactions with Aleks by assessing whether Aleks's task interpretation matched our intent and whether its deliverable addressed a research bottleneck while supporting our research question. We also assessed the methodological approach Aleks used to design each deliverable using its reasoning logs and scripts.

We considered AI support valuable in Workflow 1 if it enabled us to test whether this vineyard management dataset could support site-specific red-leaf forecasting models for prioritized red-leaf scouting. We assessed this value through retrospective reduced-labor scouting simulations comparing the efficiency of model-informed scouting against random, systematic, and adaptive scouting using independently collected full scouting records as the benchmark. We selected model development as a candidate for AI support because iterative, domain-informed model design and testing can require substantial time and expertise that may not be available for all research teams, growers, and consultants (Olson et al., 2016). Earlier evaluations showed that Aleks could autonomously develop and refine prediction models from vineyard records (Jin et al., 2025).

We considered AI support valuable in Workflow 2 if it resulted in an implementable sampling plan to compare virus detection rates between vines with higher and lower model scores. We compared the plan's design, field implementation, and virus-testing outcomes with those of an independently developed domain scientist plan. Leveraging spatial data for precision vineyard sampling can also require time and analysis that growers and researchers may not have capacity for (Song et al., 2022), which motivated us to consider AI support for designing project-specific sampling plans.

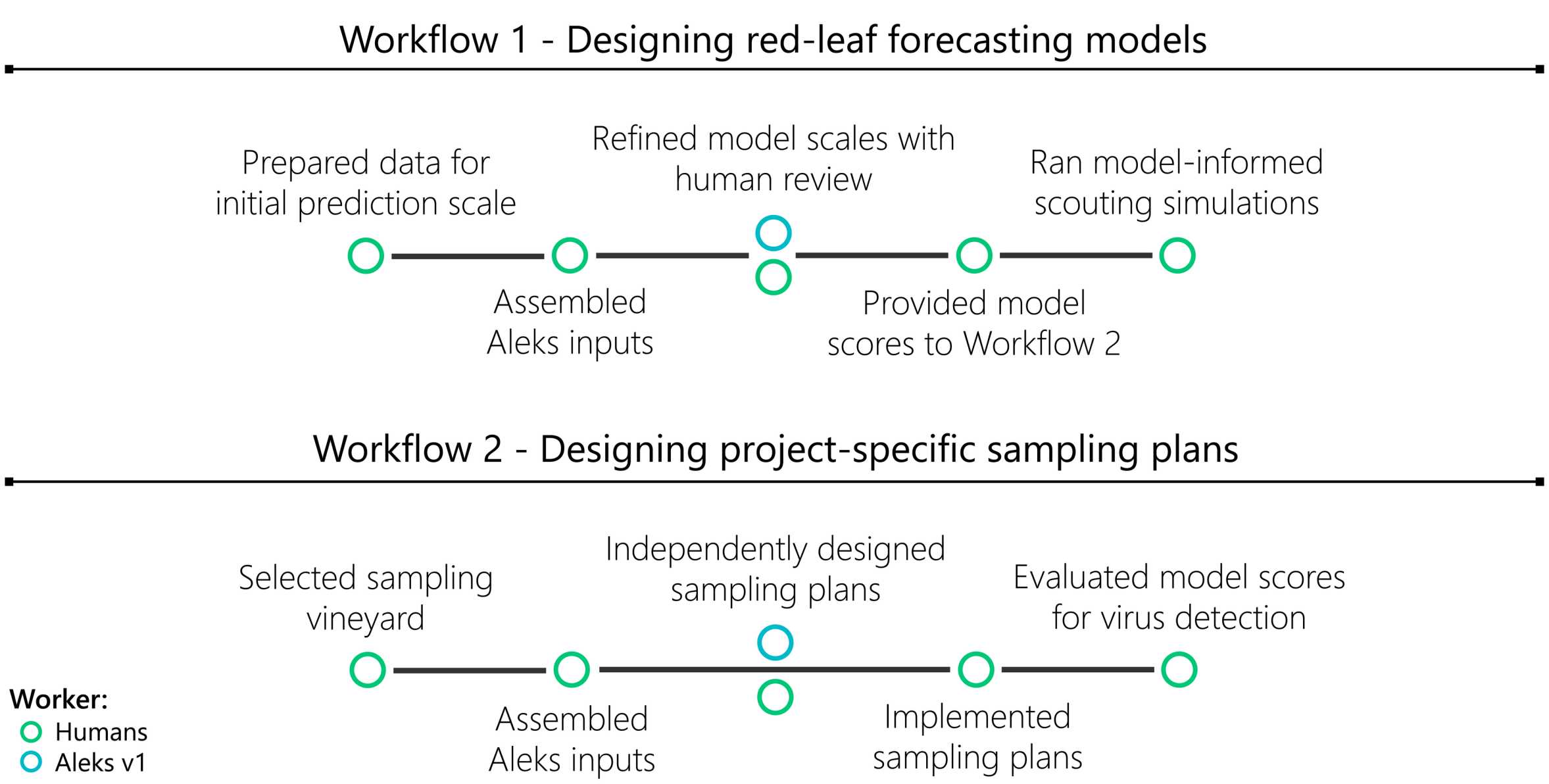

**Figure 1. Human and AI task involvement across the two field research workflows assessed in this case study.** Together, these workflows evaluated whether commercial vineyard management records and remote-sensing products could be used to prioritize vines for red-leaf scouting and virus testing. Green and blue symbols indicate human and Aleks v1 participation, respectively. We assessed AI support in Workflow 1 through a post hoc review of modeling methods and retrospective scouting simulations using independent scouting records. We assessed AI support in Workflow 2 by comparing the design, field implementation, and virus-testing outcomes of Aleks's plan with those of an independently developed domain-scientist plan.

## 2.2 – Vineyard management records and dataset preparation

The provided vineyard management records covered 296,489 vine positions across 110 commercial vineyard blocks and approximately 140 hectares in Napa Valley, California. The data contained ranch, block, row, vine coordinates, cultivar, vine spacing, annual georeferenced red-leaf and girdling observations, and repeated drone-derived remote-sensing measurements. During routine commercial scouting, vineyard management personnel walked vineyard rows and recorded scouting observations of red-leaf symptoms and vine girdling using phone GPS or VineView PinPoint devices with real-time kinematic (RTK) correction from 2021 through 2025. Remote-sensing measurements were collected from 2022 through summer 2025, with timing and coverage varying among years and blocks. We received the fall 2025 scouting records after completing 2025 model predictions and the molecular testing campaign and used those scouting records for independent model evaluation.

To prepare the vineyard management records for modeling, we generated prediction units at 10-meter, 3-meter, and vine scales by aligning the records to a common set of mapped vine positions. At the vine scale, each vine position served as one prediction unit. We then counted and assigned red-leaf and girdled-vine records within 1.5 m to each vine position for each year of data. In order to account for variations in the GPS accuracy of scouting observations, a single scouting record could contribute to the counts of more than one position. Girdling observations attributed to three-cornered alfalfa treehopper feeding (Flasco, Hoyle, et al., 2023) were available for 2023 and 2024. Supplementary Methods Section S2.2 describes details about spatial alignment, interpolation of remote-sensing measurements, and construction of multi-scale prediction units.

## 2.3 – AI system design and interaction

Aleks v1 is a multi-agent AI research system comprised of domain-scientist, data-analyst, and machine-learning-engineer agents that communicate through shared memory (Jin et al., 2025). The domain-scientist agent evaluates biological relevance using literature supplied by the research team. The data-analyst agent proposes and revises analytical approaches. The machine-learning-engineer agent generates and executes Python scripts and revises failed scripts using returned errors. Aleks used DeepSeek-V3-0324 for both workflows. Auto-sklearn ran in Python 3.8 for model development. We communicated task objectives and project context to Aleks through role prompts, task prompts, datasets and literature before each run. Once initiated, each Aleks run proceeded through its internal iterations without human intervention. Supplementary Methods Section S2.3 reproduces the role and task prompts and describes the datasets, publications, and project records supplied to Aleks for each workflow.

For Workflow 1, Aleks received the vineyard dataset and 30 publications selected by researchers and listed in File S3. The task was to use historical observations and remote-sensing measurements available before August 2024 to predict 2024 red-leaf occurrence. The prompt did not prescribe a model type, predictor set, class-imbalance treatment, or performance metric. We limited each model development run to 20 iterations, matching the budget used during development of Aleks v1 (Jin et al., 2025). File S1 contains the reasoning log of the final model development run.

For Workflow 2, Aleks received the final vine-scale model-development script (File S4), the Workflow 1 reasoning log (File S1), the 2025 vine-scale model outputs, the vineyard dataset, and 44 publications listed in File S3. The prompt asked Aleks to select 50 vines from the sampling vineyard to validate the 2025 model predictions. It did not specify our intended comparison of virus detection rates between vines with higher and lower model scores, nor did it emphasize the potential rarity of virus-positive vines. The prompt left Aleks to determine the sampling design and the criteria for assessing it. We limited the run to five iterations. File S2 contains Aleks's reasoning log for the sampling plan design run. Supplementary Methods Section S2.5.1 describes our inputs to Aleks in Workflow 2, and Aleks's reported procedures for selecting positions.

## 2.4 – Workflow 1: Model development and evaluation

### 2.4.1 – Model development and 2025 forecasting

The Aleks model reported by Jin et al. (2025) provided the starting 10-meter prediction scale for Workflow 1. Model refinement took place from our first meeting with vineyard managers on 21 August through 2 September 2025. We assessed the operational relevance of 10 m predictions *in situ* with vineyard stakeholders, then used Aleks to develop 3 m predictions as a compromise between uncertainty in recorded scouting locations and relevance to vine-level management. We ultimately selected vine-scale prediction units for final model development because scouting, molecular sampling, and vineyard management decisions occurred on individual vines.

We reviewed Aleks's reasoning log (File S1) and final model-development script (File S4) post hoc to assess its modeling choices against peer-reviewed literature. The final iteration produced an Auto-sklearn ensemble model that assigned each mapped vine position a continuous model score from 0 to 1, with higher model scores indicating greater model support for at least one 2024 red-leaf record. Positions with model scores of 0.50 or greater were classified by Aleks as predicted positives and those with lower model scores as predicted negatives. Supplementary Methods Section S2.4.1 documents predictor construction, preprocessing, class-imbalance treatment, model-fitting settings, and internal evaluation of the final model.

We preserved Aleks's original modeling approach and final model parameters to assess the operational value of its deliverable. On 2 September 2025, we applied the model to updated 2025 inputs to generate one harvest-2025 model score for each of the 296,489 vine positions. Table S4 lists the input substitutions, and Supplementary Methods Section S2.4.2 describes predictor preparation.

### 2.4.2 – Validation of 2025 red-leaf forecasts using independent scouting

The vineyard management team conducted 2025 visual scouting of the vineyard management area from September through December 2025 without access to model predictions. We received the completed scouting records in February 2026 after the 2025 model scores had been generated. The "New in 2025" evaluation included vine positions without a red-leaf record from 2021 through 2024. The "All in 2025" evaluation included all vine positions. In both evaluations, a 2025 red-leaf record defined an observed positive, and the remaining eligible positions were observed negatives.

We calculated vine-scale model-performance metrics across the full management area and within the sampling vineyard described in Section 2.5. ROC AUC and average precision summarized how well model scores ranked observed-positive positions above observed-negative positions. Average-precision enrichment was calculated as average precision divided by observed-positive prevalence. For the full management area, we calculated block-adjusted ROC AUC to assess ranking within blocks, following Janes et al. (2009). Precision and recall summarized classification at the 0.50 threshold. Results are reported in Tables 1 and S1. Supplementary Methods Section S2.4.3 provides metric definitions, calculation procedures, and software details.

### 2.4.3 – Whole-row scouting simulations

Using vine-scale model scores and 2025 scouting results, we retrospectively compared five whole-row scouting strategies across the full management area through 45% survey effort. Survey effort was the percentage of all vine positions contained within surveyed rows, and recall was the percentage of vine positions with newly recorded 2025 red-leaf observations contained within those rows. Random scouting ordered rows randomly, and systematic scouting surveyed alternating rows. We adapted the sampling approach of Rohrs et al. (2025) to whole-row visual scouting: Adaptive scouting initially surveyed every third row and expanded around these rows by two rows in each direction when they contained a 2025 red-leaf observation. Model-prioritized scouting ranked rows by the mean score of their highest-scoring 10% of vine positions, while model-adaptive scouting combined this ranking with adaptive expansion. Figure 3C illustrates row prioritization in the sampling vineyard.

Random, systematic, and adaptive scouting were each evaluated across 5,000 simulated routes, while the two model-based strategies each produced one route. A route was the ordered sequence of rows surveyed. For strategies with multiple routes, the 2.5th and 97.5th percentiles summarized variation in recall among routes (Figure 2). We also compared strategies with dataset-wide random scouting and within-block random row selections (Figure S1). Recall exceedance was the percentage of comparisons in which a strategy achieved strictly greater recall than its comparator. Supplementary Methods Section S2.4.4 describes how we constructed scouting routes, applied adaptive expansion, and conducted dataset-wide and within-block random comparisons.

## 2.5 – Workflow 2: Sampling plan design and evaluation

### 2.5.1 – Sampling plan development

Both the domain-scientist and Aleks sampling plans were developed from 2025 vine-scale model scores for a 5.1-hectare Cabernet Sauvignon vineyard (sampling vineyard) within the 140-hectare management area. The sampling vineyard contained 12,792 candidate vine positions across 78 rows and two blocks. The vineyard had four years of historical vineyard records, approximately 1.8-m spacing between rows and vines, and bilateral-cordon training. The vineyard contained a model score gradient from a historically symptomatic northeastern region toward a southwestern region with lower model scores (Figure 3A–B). The independently developed domain-scientist and Aleks plans aimed to sample 100 and 50 vine positions, respectively.

After receiving the sampling plan prompt, Aleks generated Python code to select vine positions and used summaries of model scores, spatial coverage, and pairwise distances to assess and revise its designs. The design accepted by Aleks targeted 10 positions from each of five model score quintiles and a minimum spacing of 15 meters. The Aleks plan lacked replacement instructions for empty or unsuitable vine positions. We received the reasoning log and final selected positions from Aleks.

Given a limited testing budget and the rarity of red-leaf vines in historical records, our sampling plan oversampled vines with higher model scores to increase the chance of obtaining virus-positive observations if infection was more frequent at higher model scores (Parnell et al., 2014). We divided candidate positions into four subclasses based on whether their model score was below 0.50 or at least 0.50 and whether they were within 3 m of a 2024 red-leaf observation. We selected 25 positions from each subclass, excluding positions with 2024 red-leaf records. Supplementary Methods Section S2.5.1 further details vineyard and vine selection, plan design, and Aleks's reported plan-development procedures.

### 2.5.2 – Field implementation and virus testing

The vine positions selected for sampling in the sampling vineyard were located using row and vine counts, with supporting grid alignment provided by a VineView PinPoint RTK device. In September 2025, we collected petiole tissue from 97 domain-scientist-plan vines. During September petiole sampling, we selected replacements for plan positions that were empty or contained vines younger than three years, following the rationale described in Supplementary Methods Section S2.5.2. Between the 9$^{th}$ and 10$^{th}$ of December 2025, we collected cane tissue from the same six canes used for petiole collection at those 97 vines and from six canes at 41 Aleks-plan positions (Figure 3D). These tissue and timing choices were consistent with studies supporting late-season petiole and dormant cane testing for GRBV and GLRaV-3 (DeShields & Kc, 2023; Osman et al., 2018; Setiono et al., 2018; Tsai et al., 2012). The same researcher implemented the domain-scientist petiole-sampling plan and the Aleks cane-sampling plan. Tissue from the six canes was pooled into one cane sample per vine. AgriAnalysis (Davis, California, USA) tested petiole and cane samples for GRBV by reverse transcription polymerase chain reaction (RT-PCR) and GLRaV-3 by enzyme-linked immunosorbent assay (ELISA). Laboratory staff received sample identifiers but not model scores. Supplementary Methods Section S2.5.2 provides tissue-collection and handling details.

**2.5.3 – Sampling plan and model score evaluation**

We compared the positions selected by each plan using model score distributions, proximity to 2024 red-leaf observations, minimum spacing, and vineyard rows represented. We evaluated field implementation using the percentage of planned positions tested, the median count of tested vines per sampled row, and counts of GRBV-positive and GRBV-negative cane samples. Table 2 reports these comparisons. Supplementary Methods Section S2.5.3 explains how we defined calculations for comparing sampling plans, calculated metrics for reporting petiole and cane test results, and evaluated model scores against virus detection.

For the 97 domain-scientist-plan vines, we calculated the percentage with matching petiole and cane test classifications separately for GRBV and GLRaV-3. We used December cane results for subsequent model score analyses, as cane tissue was collected under both plans. One vine selected by both plans contributed to each plan's summary but was counted once in the combined sample of 137 vines. We calculated recall as the percentage of GRBV-positive vines with model scores ≥0.50. We also summarized GRBV detection percentages among vines with scores ≥0.50 and <0.50, median model scores by GRBV status, and ROC AUC (Table 3). Analyses were conducted in R 4.5.2.

# 3 – Results

## 3.1 – Workflow 1: Designing and evaluating red-leaf forecasting models

### 3.1.1 – Aleks designed red-leaf forecasting models at multiple prediction scales

Aleks developed models for 10 m × 10 m and 3 m × 3 m grid cells and individual vine positions. We iteratively assessed the suitability of these prediction scales for field use with vineyard stakeholders. We selected vine-scale prediction units to match the scale of scouting, sampling for virus testing, and vineyard management of this specific management area.

Aleks completed 20 vine-scale model-design iterations in 145 minutes using data from all 296,489 vine positions. It evaluated red-leaf counts and red-leaf presence or absence as alternative potential outcomes. Aleks also evaluated combinations of original and derived predictors and tested class weighting and the synthetic minority oversampling technique (SMOTE). Aleks used Auto-sklearn to evaluate tree-based ensembles, nearest-neighbor models, linear classifiers, discriminant analysis, and naive Bayes models. Candidate predictors included historical red-leaf observations, canopy structure, vegetation indices, vine presence, coordinates, cultivar, and vine spacing. Derived predictors included annual changes in enhanced vegetation index (EVI) and canopy area, mean historical red-leaf counts among each vine's five nearest neighbors, a sum of 2021–2023 red-leaf counts weighted toward more recent observations, and interactions between EVI and canopy measurements (File S1). Our post hoc review identified peer-reviewed precedent for Aleks's temporal EVI difference and nearest-neighbor averaging as predictors (Supplementary Methods Section S2.4.1).

The final vine-scale model used 16 predictors to classify 2024 red-leaf presence (Table S4). Aleks reported 97% accuracy on its nonindependent internal evaluation subset (File S1). Our post hoc review found that Aleks calculated imputation medians and applied SMOTE before the training/evaluation split, allowing records later assigned to evaluation to influence preprocessing (File S4; Blagus and Lusa, 2015; Perez-Lebel et al., 2022). The evaluation subset also contained synthetic records and was labeled 50% red-leaf positive, compared with 3.0% in the original dataset. The reported internal estimates therefore did not establish performance on independent observations with the original class distribution. SMOTE also preceded Auto-sklearn's internal validation, potentially affecting model and hyperparameter selection. We applied the final model to updated 2025 inputs, generating one harvest-2025 model score for each of the 296,489 mapped vine positions (Section 2.4.1).

#### 3.1.2 – Vine-scale model scores distinguished 2025 red-leaf observations

In the "New in 2025" evaluation, model scores distinguished observed-positive from observed-negative positions more strongly across the management area than within blocks. Dataset-wide ROC AUC was 0.659, compared with a block-adjusted ROC AUC of 0.516. Average precision was 0.65%, 1.89 times the 0.35% observed-positive rate. The "All in 2025" evaluation showed the same pattern of stronger dataset-wide discrimination (Table 1). At the 0.50 model score threshold, 32 of the 869 observed-positive positions in "New in 2025" were classified positive, corresponding to 3.68% recall (Table 1). Among the 2,223 positions classified positive using the 0.50 threshold, 1.44% were observed positive (Table S1). Within the sampling vineyard, ROC AUC was 0.613 for "New in 2025" and 0.615 for "All in 2025" (Table S1).

**Table 1. Vine-scale model performance for predicting 2025 red-leaf observations across the full 140-ha management area.** "All in 2025" evaluates all 296,489 vine positions. "New in 2025" evaluates the 251,749 positions without a previous red-leaf record. Table S1 reports expanded metrics addressing vine-scale model performance.

| | **All in 2025** | **New in 2025** |
|---:|:---:|:---:|
| Observed + positions, n | 1,163 | 869 |
| Observed + prevalence | 0.39% | 0.35% |
| Average precision | 0.70% | 0.65% |
| Average precision enrichment (× prevalence) | 1.79x | 1.89x |
| Median scores, observed + / - | 0.263/0.197 | 0.260/0.187 |
| Dataset-wide ROC AUC | 0.667 | 0.659 |
| Block-adjusted ROC AUC | 0.572 | 0.516 |
| Recall at 0.50 threshold, n/N | 45/1,163 (3.87%) | 32/869 (3.68%) |

#### 3.1.3 – Vineyard management data enabled scouting comparisons

At 45% survey effort in our retrospective whole-row scouting simulations, model-adaptive scouting achieved 94.13% recall, compared with mean recall of 85.85% for adaptive and 45.02% for random scouting (Figure 2; Table S2). Model-prioritized scouting achieved 89.76% recall, while mean recall for systematic scouting was 44.86%. Both model-based strategies exceeded every dataset-wide random route at 10%, 20%, and 45% survey effort. Adaptive routes exceeded their paired random routes in an increasing percentage of comparisons as effort increased, whereas systematic scouting remained near random (Table S2).

Model-adaptive, model-prioritized, and adaptive scouting had larger recall advantages over dataset-wide random selection than over within-block random selection (Table S3; Figure S1). At 10% survey effort, both model-based strategies had lower recall than the within-block random mean. At 45% effort, mean recall advantages were 2.48 percentage points for model-adaptive, 2.91 for model-prioritized, and 4.65 for adaptive scouting. By 45% effort, model-adaptive, model-prioritized, and adaptive scouting exceeded within-block random recall in nearly all matched comparisons, whereas systematic remained near random. These comparisons indicated smaller, effort-dependent gains in recall from row selection within blocks for adaptive and model-adaptive scouting.

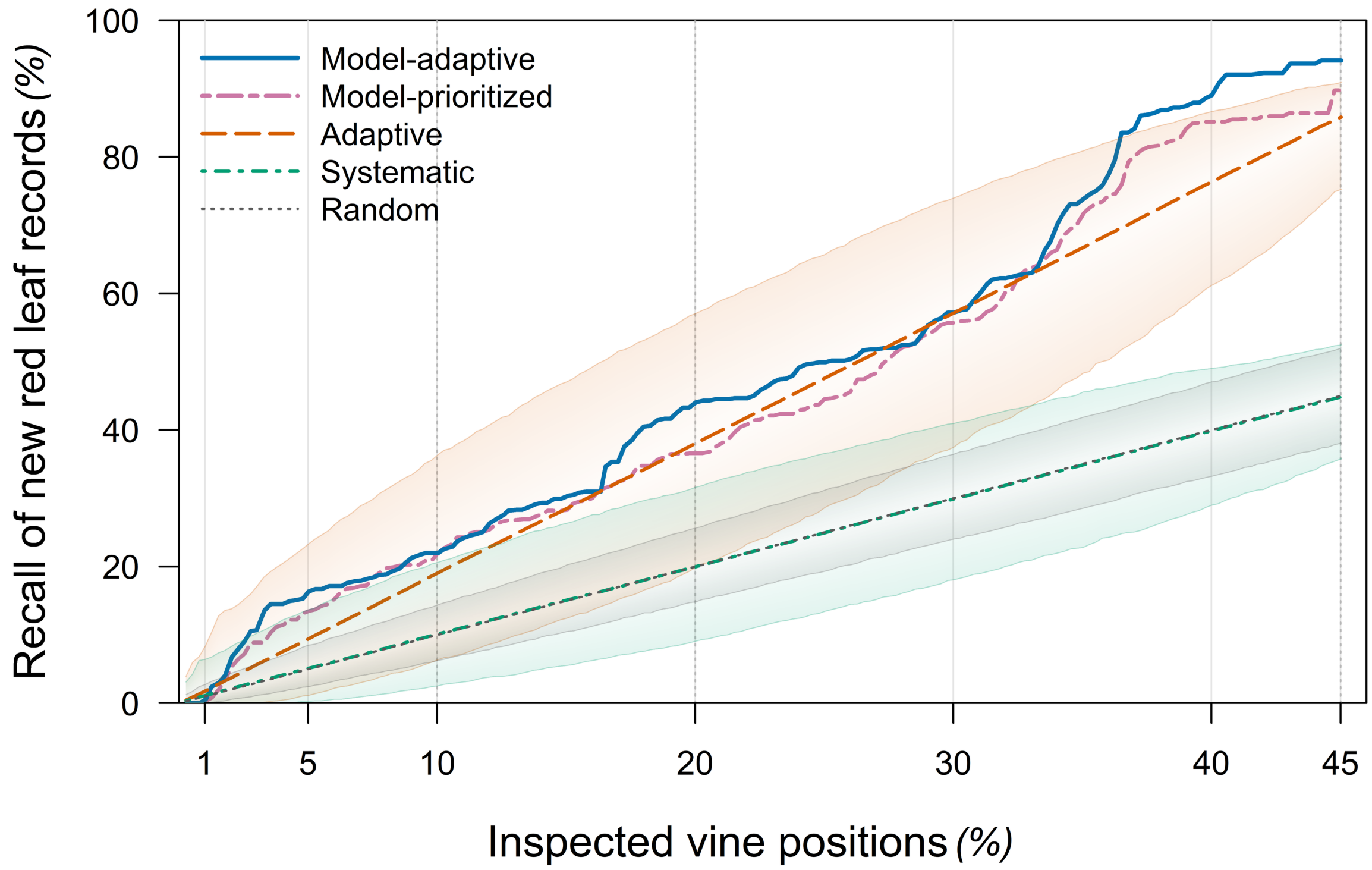

**Figure 2. Recall of vine positions with red-leaf observations newly recorded in 2025 across survey effort for five whole-row scouting strategies in the full 140-hectare management area.** For dataset-wide random, systematic, and adaptive scouting, lines show mean recall and ribbons span the 2.5th to 97.5th percentiles across 5,000 simulated routes. Model-prioritized and model-adaptive lines each show recall from one deterministic route. Survey effort was the percentage of all vine positions contained within surveyed rows; dotted vertical guides mark 10%, 20%, and 45% survey effort, which are summarized in Table S2.

## 3.2 – Workflow 2: Designing and evaluating project-specific sampling plans

### 3.2.1 – AI and domain-scientist plans established distinct priorities

Aleks evaluated four sampling designs across five iterations and produced a 50-position sampling plan for the sampling vineyard in 15 minutes (File S2). The final Aleks plan emphasized model score coverage and spatial dispersion, whereas the domain-scientist plan emphasized the highest model score quintile, proximity to 2024 red-leaf observations, and labor-efficient sampling. Compared with the Aleks plan, the domain-scientist plan included more positions with scores ≥0.50 (50.0% versus 10.0%) and within 3 m of a 2024 red-leaf observation (50.0% versus 20.0%) but represented fewer vineyard rows (10 versus 36 of 78; Table 2). The domain-scientist plan placed 64.0% of positions in the highest sampling-vineyard score quintile, whereas the Aleks plan placed 12.0 to 26.0% in each quintile. The final Aleks plan had a minimum spacing of 5.5 m, below the 15 m minimum Aleks had proposed. Aleks noted this discrepancy in the fifth iteration but accepted the plan (File S2). Aleks's sampling design script was not included in its deliverables, preventing reconstruction of the exact selection procedure.

December 2025 cane testing covered 97 vines under the domain-scientist plan and 41 under the Aleks plan, fulfilling 97.0% and 82.0% of the respective 100- and 50-vine sampling targets (Figure 3D; Table 2). Three domain-scientist targets remained unfilled after *in situ* replacements during September petiole sampling, and we did not add replacements for them in December. The Aleks plan provided no replacement instructions for missing or young vines, and we were unable to sample nine Aleks-selected positions. The median count of tested vines per sampled row was 10 for the domain-scientist plan and one for the Aleks plan (Table 2). Property owners requested removal of all vines from the sampling vineyard in winter 2026, after field sampling was completed between the 9th and 10th of December 2025, which was not anticipated by any party at the project's outset.

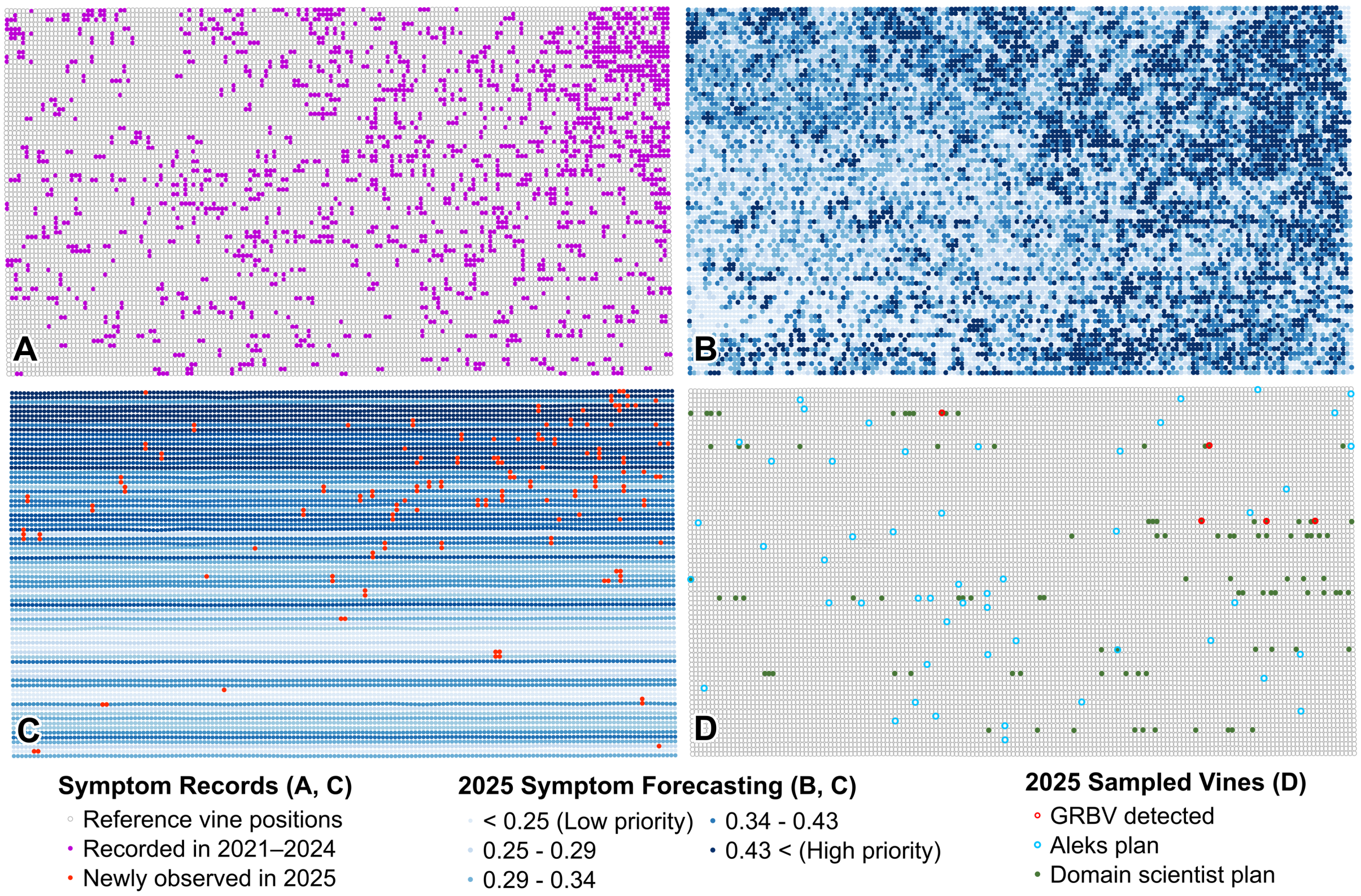

**Figure 3. Red-leaf symptom observations, model forecasts, and virus-testing outcomes in the 5.1-hectare sampling vineyard.** Points represent mapped vine positions. (A) Red-leaf observations spanning 2021–2024. (B) Vine-scale model scores for forecasting 2025 red-leaf symptom observations derived from scouting history, remote-sensing measurements, and vineyard metadata. Darker blue indicates higher scores. (C) Model-score-based row prioritization derived from vine-scale forecasts in panel B; darker blue indicates higher priority rows. Red points mark newly recorded red-leaf observations in 2025. (D) Sampled positions under the domain scientist (green) and Aleks v1 (cyan) sampling plans, for assessing associations between model scores and virus detection. Red rings mark vines positive for grapevine red blotch virus (GRBV) from December 2025 cane testing with RT-PCR.

**Table 2. Planned and realized sampling of domain-scientist and Aleks sampling plans in the sampling vineyard.** Planned threshold + denotes planned sampling positions with model scores ≥0.50. Model score quintile values are percentages of planned positions; approximately 20% per quintile would match the vineyard's model score distribution. Minimum spacing is the smallest distance between planned positions; row representation is reported relative to the sampling vineyard's 78 rows. Plan completion reports tested vines, including replacements, relative to each plan's target sample size. Grapevine red blotch virus (GRBV) RT-PCR results and median tested vines per sampled row use December 2025 cane samples.

| | **Domain-scientist plan** | **Aleks plan** |
|---:|:---:|:---:|
| Planned threshold +, n/N | 50/100 (50.0%) | 5/50 (10.0%) |
| Planned within 3 m of 2024 red-leaf observations, n/N | 50/100 (50.0%) | 10/50 (20.0%) |
| Planned model score quintiles, Q1–Q5 (%) | 7.0 / 9.0 / 10.0 / 10.0 / 64.0 | 22.0 / 26.0 / 24.0 / 12.0 / 16.0 |
| Minimum spacing, m | 1.8 | 5.5 |
| Rows represented in the plan, n/N | 10/78 | 36/78 |
| Plan completion, n/N | 97/100 (97.0%) | 41/50 (82.0%) |
| Median tested vines per sampled row | 10 | 1 |
| GRBV +/- cane samples, n | 5/92 | 0/41 |

### 3.2.2 – GRBV-positive vines had higher model scores in 2025

Across the two plans, 138 plan-specific cane-test records represented 137 unique vines, with one GRBV-negative vine selected by both plans. Petiole and cane test results agreed for GRBV in all 97 vines with paired samples and for GLRaV-3 in 96 of 97. One vine was GLRaV-3-positive in petiole tissue, but repeated testing of its cane sample returned negative GLRaV-3 results. No GLRaV-3 was detected in cane samples.

Five vines were GRBV-positive, all selected by the domain-scientist plan, and four had model scores ≥0.50. All five GRBV-positive samples came from vine positions without recorded red-leaf symptoms from 2021 through 2025. Within the selected sample, GRBV was detected in 8.16% of vines with scores ≥0.50 and 1.14% of vines with scores <0.50. GRBV-positive vines also had higher median scores than GRBV-negative vines (0.525 versus 0.407), with ROC AUC 0.735 (Table 3).

**Table 3. Red-leaf model scores and grapevine red blotch virus (GRBV) RT-PCR cane test results for 137 vines selected through domain-scientist and Aleks sampling plans in the sampling vineyard in December 2025.** ROC AUC summarizes score ranking among the sampled vines. Threshold + and threshold - denote model scores ≥0.50 and <0.50, respectively. GRBV detection percentages are followed by positive/tested counts. Sampled quintile values represent unique tested vines; approximately 20% per quintile would match the vineyard's model score distribution.

| | **Result** |
|---:|:---:|
| Tested vines, n (GRBV +) | 137 (5) |
| Median model score, GRBV + / GRBV - | 0.525 / 0.407 |
| ROC AUC | 0.735 |
| GRBV + recall at 0.50 threshold, n/N (%) | 4/5 (80.00%) |
| GRBV detection percentage, threshold + / threshold - | 8.16% (4/49) / 1.14% (1/88) |
| Sampled model score quintiles, Q1–Q5 (%) | 10.2 / 14.6 / 17.5 / 10.2 / 47.4 |

# 4 – Discussion

## 4.1 – Vineyard records enabled site-specific decision support modeling

This case study demonstrates that vineyard management records and remote-sensing products can inform site-specific red-leaf forecasting models. These records can also enable comparisons between scouting methods, extending their use from characterizing disease patterns and management practices (Flasco et al., 2025; MacDonald et al., 2021) to supporting scouting decisions under resource constraints. With this management dataset, retrospective scouting simulations indicated that model forecasts were more useful for allocating scouting effort among blocks than for prioritizing rows within blocks in 2025. Adaptive row scouting also achieved high recall without model scores, suggesting a potentially useful approach for resource-limited scouting where forecasting models are unavailable. The performance of model-guided and adaptive scouting at other sites or with other datasets remains to be evaluated. Resource prioritization based on model scores may be particularly useful when labor constraints prevent reliable seasonal coverage of entire management areas. Model scores could support designating vineyard areas for intensive or sparser human scouting, or allocating coverage among human, robotic, and airborne scouting (Laroche-Pinel, Cooper, et al., 2025; E. Liu et al., 2026; Y. Liu et al., 2025; Wang & Pagay, 2024). To address the possibility that consistently limited scouting could create persistent gaps in red-leaf records, future studies could test whether rotating reduced scouting between vineyard areas can limit the downstream impacts of observational gaps. Comparing symptom detections and costs between these approaches and current grower practices would help assess their potential for commercial decision support.

While the model forecasts in this case study were not as valuable for within-block red-leaf forecasting, the vineyard management records used here were not originally intended for vine-scale decision support modeling. Future studies should test whether vineyard datasets collected specifically for forecasting can improve model performance for within-block scouting prioritization. Tracking virus diagnoses, nutrient deficiencies, irrigation, roguing, red-leaf timing and persistence, and other potentially relevant features at vine positions located with RTK GPS may help models distinguish local red-leaf patterns associated with disease, abiotic stress, or other biotic damage (Alisaac et al., 2026; Cieniewicz & Fuchs, 2025). The vineyard features described above may also help spectral models differentiate virus-associated spectral changes (Laroche-Pinel, Singh, et al., 2025; Romero Galvan et al., 2023; Wang & Pagay, 2024) by accounting for biotic and abiotic influences on vine reflectance that can confound disease detection (Alisaac et al., 2026; Laroche-Pinel et al., 2024).

The tentative association between higher model scores and GRBV detection in our 137-vine sample, which included five positive vines, motivates evaluating whether model-score-guided sampling could increase the number of newly identified virus-positive vines per test compared with current sampling methods. For example, future studies could compare adaptive cluster sampling initialized using model-score-guided versus random vine selection within sections of vineyard blocks (Fendell-Hummel & Cooper, 2025; Rohrs et al., 2025). Researchers could also test whether adding the features proposed for improved red-leaf scouting, specifically longitudinal molecular-testing and roguing records, helps model scores distinguish vines with positive versus negative virus-test results. Future validation of model-score-guided testing for GRBV or GLRaV-3 should account for assay choice, sampled tissue, and sampling date in reported results, given that negative tests do not necessarily establish a lack of infection (Constable et al., 2012; Crespo-Martínez et al., 2023; Klaassen & Al Rwahnih, 2026; Setiono et al., 2018).

Given the investment required to collect detailed, multiyear vineyard records, regionally trained red-leaf forecasting models may support local scouting and virus-testing decisions after adaptation with targeted site-specific training data (Bueechi et al., 2025). To identify which local inputs and training scales best support local forecasting, future work could compare forecasting performance among management-area-specific models, regional models without local adaptation, and regional models adapted using different sets of local predictors (Bueechi et al., 2025; Colaço et al., 2025). To limit sharing of local management records, researchers could test federated training for regional model development (Li et al., 2024). Researchers could also test whether combining site-specific and regional model scores improves local forecasting (Breiman, 1996; Loewinger et al., 2024).

### 4.2 – The value of AI support differed between workflows

In this case study, our two human-AI interactions produced modeling and sampling-design deliverables that varied in their support of our project goals. Our human-AI collaboration in Workflow 1 extends examples of AI support in research from laboratory studies (Ghareeb et al., 2026) to live agricultural field research, where a bounded AI task with iterative human review addressed our modeling throughput bottleneck while supporting our research questions. Despite the preprocessing and reporting errors identified during post hoc review of Aleks's final 2024 model design, model-informed scouting improved recall in retrospective simulations. These results demonstrated the potential of vineyard management records and remote-sensing products for site-specific scouting decision support, addressing our research question. Additionally, AI-supported model design directly contributed to the throughput of our project by allowing our research team to focus on preparing datasets for multiple prediction scales and planning for molecular sampling. If slower *in situ* project design or sampling had delayed December 2025 cane collection until after the unexpected removal of the sampling vineyard in winter 2026, we would not have been able to compare the field outcomes of the domain-scientist and AI plans, or the virus-test results from petiole and cane tissue.

Researchers developing a grower-ready system for site-specific red-leaf forecasting could further assess dataset preparation and scouting simulations as workflow tasks for AI automation. AI systems completing these tasks may benefit from iterative human review and separate evaluation of each task, particularly given the documented limitations of AI-driven data loading and processing (Z. Chen et al., 2025). To address Aleks's modeling mistakes in Workflow 1, researchers could test whether supplying Aleks with peer-reviewed literature covering these mistakes helps it independently avoid them in the future (Blagus & Lusa, 2015; Krstajic et al., 2014). While iterative review of preprocessing and modeling scripts could address modeling errors (Ifargan et al., 2025), the time and expertise required would have reduced the value of AI support in this project.

Workflow 2 revealed gaps in alignment between Aleks's independently designed sampling plan and the domain scientists' operational objectives. We tested whether Aleks could independently infer our sampling objective and field requirements from the supplied inputs, including the Workflow 1 reasoning log and final model predictions, the vine-scale vineyard management dataset spanning 2021-2024, and literature discussing roguing and vineyard-scale decision support. According to its reasoning logs and selected vine positions, Aleks aimed to sample evenly across model score quintiles and spread selected vines throughout the vineyard. The domain-scientist plan instead emphasized vines with higher model scores to increase opportunities for virus detection under our limited testing budget. Field implementation revealed that Aleks neither requested site-specific roguing information nor addressed the possibility of missing target vines in its sampling plan, and its selected positions spanned more than three times as many vineyard rows as ours. Although Aleks's plan contributed negative virus-test results to help associate model scores with virus detection, its incomplete implementation and dispersed sampling positions limited its operational value in this project.

Future workflow trials could test whether the operational value of Aleks's sampling designs improves with an explicit explanation of downstream sampling objectives and an introduction of vineyard-specific vine-status information. A future Aleks design could also support requests for clarification during plan development, similar to the interactive plan refinement used in PhenoAssistant (F. Chen et al., 2026). The missing Workflow 2 script prevented us from reconstructing Aleks's exact selection procedure, supporting a requirement that future Aleks deliverables include executable scripts, reasoning logs, task inputs, and software versions needed for complete task review and reproduction (Sandve et al., 2013).

### 4.3 – Live workflow evaluations can guide the development of reliable AI support

This case study supports evaluating human-AI interactions through live research workflows to assess how project conditions shape the value of AI deliverables and how those deliverables help address research questions. Linking our review of deliverable design to field implementation and project outcomes allowed us to assess the methodological quality of AI support alongside its operational value. The workflow-scale evaluation of our two live human-AI interactions demonstrated how some methodological oversights by AI do not necessarily prevent useful research support, while sound methodological precedent alone does not ensure that an AI deliverable will provide operational value.

This case study also illustrates how project objectives and available resources help determine which approaches are suitable for evaluating a live human-AI interaction. In Workflow 2, comparing Aleks's sampling plan to ours enabled a quantitative comparison of sampling outcomes and a qualitative assessment of whether Aleks's objective interpretation matched our intent. To protect Workflow 2's objective of assessing whether higher model scores were associated with more frequent virus detection, we allocated more of our limited testing resources to our plan, reducing our dependence on an AI interaction protocol not yet shown to produce an implementable deliverable. In Workflow 1, we lacked the time, resources, and labor to develop and evaluate a parallel human comparator for AI-enabled red-leaf forecasting model design. Earlier modeling results (Jin et al., 2025) supported using our existing workflow for red-leaf model development, with human review incorporating site-specific context as needed. Therefore, we chose to use Aleks to generate our red-leaf forecasts to best support our constraints around planning and executing the field project within the year. In the absence of a direct human comparator, we evaluated the human-AI interaction in Workflow 1 by assessing Aleks's final forecasting model against independent 2025 scouting records and auditing its modeling decisions against peer-reviewed literature. In this way, researchers conducting live workflow assessments can balance the resources devoted to evaluating AI support with those needed to meet the project's research objectives.

Repeated workflow evaluations could identify human-AI interaction protocols that reliably address recurring bottlenecks for field research teams. Building on our results from Workflow 1, researchers using site-specific data for scouting decision support could test whether changes to human-AI interactions improve the recall of model-informed scouting relative to other methods. In the case of AI-supported sampling plan design, researchers could vary interaction protocols and compare how well human and AI plans address project-specific sampling constraints, such as the rarity of red-leaf observations in Workflow 2. These evaluations could also test whether custom AI systems provide greater operational value than off-the-shelf systems for particular research tasks (Gottweis et al., 2026). To develop reliable interaction protocols for live trials, field researchers can consider approaches to AI interaction that have worked elsewhere, such as in plant phenotyping (F. Chen et al., 2026), biomedicine (Ghareeb et al., 2026; Gottweis et al., 2026), and agronomy (Chamara et al., 2026). Researchers could also use existing resources for interaction design (Amershi et al., 2019) and workflow design (Microsoft, 2026) to develop reusable and shareable workflow templates. Comparisons across AI systems, interaction protocols, and project contexts can help researchers design human-AI workflows that provide consistent value during live field research.

## Conclusions

This case study demonstrates how live workflow-level evaluations can be used to assess the value of AI support in field research. In Workflow 1, AI-supported model development addressed a throughput bottleneck and helped us assess the potential of longitudinal vineyard records and remote-sensing products for site-specific virus decision support. In Workflow 2, we compared virus detection rates across model scores, but Aleks's sampling plan was only partially implementable and did not account for the potential rarity of virus-positive vines, limiting its operational value. These findings support assessing AI interactions by both the methodological quality of their deliverables and how well those deliverables advance research objectives under project-specific constraints. Repeated workflow evaluations across projects could help identify interaction protocols which reliably support agricultural field research.

## Author contributions (CRediT)

**Niko Carvajal Janke.** Conceptualization; Methodology; Formal analysis; Investigation; Data curation; Writing - original draft; Writing - review & editing. **Daoyuan Jin.** Methodology; Software; Investigation; Formal analysis; Writing - review & editing. **Shivranjani Baruah.** Methodology; Writing - review & editing. **Nicholas Gunner.** Data curation; Writing - review & editing. **Jacob Maus.** Methodology; Resources. **Yu Jiang.** Conceptualization; Methodology. **Kaitlin M. Gold.** Conceptualization; Methodology; Writing - review & editing.

## Acknowledgements

We thank the Napa Valley wine industry stakeholders who contributed datasets, time, and perspectives to this project.

## Data availability

Supporting files S1–S4: https://doi.org/10.5281/zenodo.23142108

Vineyard management records, vine coordinates, intermediate spatial layers, assembled datasets, the fitted model object, and additional workflow records are not publicly available because they contain or were derived from identifiable commercial-vineyard locations and proprietary management information. Controlled access to these materials may be considered upon reasonable request to the corresponding author, subject to approval by the commercial data owner and an appropriate data-use agreement.

## Funding

This work was supported in part by the NASA Acres Consortium (award 80NSSC23M0034), Cornell AgriTech, the Cornell CALS Moonshot Seed Grant Program, and the USDA NIFA Specialty Crop Research Initiative (VitisGen3, award 2022-51181-38240). The funders had no role in study design, data collection, analysis or interpretation, manuscript preparation, or the decision to submit.

## Declaration of generative AI and AI-assisted technologies in the manuscript preparation process

We used ChatGPT and Codex (OpenAI) to assist with manuscript preparation. We reviewed and revised AI-assisted code and take full responsibility for the content of this manuscript.

# Supplemental Methods

## S2.2 – Vineyard management records and dataset preparation

To create the datasets used in Workflow 1, we established a common set of mapped vine positions and used those positions as the spatial reference for integrating data from different sources and years. We then estimated remote-sensing values at these reference positions, linked vineyard scouting observations to them, and aggregated the resulting vine-scale records into coarser prediction units.

To establish the reference vine positions, we combined mapped vineyard-block boundaries with vine positions contained in VineView PureVine remote-sensing datasets. For each vineyard block, we selected the most recent PureVine dataset associated with its ranch or location, verified that its vine positions overlapped the mapped block, and retained the positions within the block boundary. Each retained vine record included plant and row identifiers, vine spacing, ranch, block, cultivar, and coordinates.

VineView supplied each PureVine acquisition as an individual-vine-point shapefile derived from drone imagery. Each file contained vine coordinates and associated remote-sensing measurements. Because the coordinates varied slightly among acquisitions, the measurement points did not always coincide with the vine positions from the reference PureVine layer. We therefore used inverse-distance weighting to estimate enhanced vegetation index (EVI), canopy area, canopy width, and vine presence at each reference position for each acquisition. Each estimate used up to the 10 nearest source points within 3 m and an inverse-distance power of 2. Vine presence indicated whether the remote-sensing product detected grapevine canopy. Measurements that could not be estimated for a given vine and acquisition date remained missing.

Vineyard management personnel used VineView and Every.Farm (formerly 'My Efficient Vineyard') applications to record red-leaf locations. Coordinates were recorded using phone GPS without RTK correction or VineView PinPoint devices with RTK correction. To add vineyard scouting records, we linked each red-leaf and girdling observation to every vine position within 1.5 m. We selected the 1.5 m radius because adjacent vines were approximately 1.8 m apart and recorded scouting locations did not always align exactly with the mapped vine positions. The search was performed independently for each vine, so a single scouting observation could be linked to more than one vine. For each mapped vine position and year, we separately counted the linked red-leaf and girdled-vine records. Vines with no linked red-leaf observations in a year received a count of zero. Depending on the year, this procedure linked 46.4% to 94.1% of red-leaf observations and 77.7% to 78.1% of girdling observations within mapped blocks to at least one vine position. Median distances between linked observations and vine positions ranged from 0.72 to 0.91 m.

To create the 10 m and 3 m model-development datasets, we grouped mapped vine positions into 3 m × 3 m and 10 m × 10 m prediction units. Within each prediction unit, we summed red-leaf and girdled-vine record counts for each year and averaged the remote-sensing values estimated at each vine position within the prediction unit. Both grids used UTM Zone 10N and originated at the minimum easting and northing of the vine-scale dataset. Prediction units without mapped vine positions were excluded.

Distance-based dataset preparation used UTM Zone 10N (EPSG 32610), and coordinates were exported in WGS84 (EPSG 4326). Scouting records were managed in VineView, QGIS, and Every.Farm. We assembled and spatially processed the datasets using Python 3.8, Every.Farm, and QGIS 3.34.15.

## S2.3 – AI system configuration and prompting

To prepare Aleks for its assigned workflow tasks, domain scientists developed task-specific prompts, supplied relevant publications as PDF files, and identified the datasets and prior project records needed for each task. The role prompt defined the expertise Aleks was asked to apply, while the task prompt described the objective and relevant project context in plain English. These materials and a fixed iteration limit were supplied to Aleks before each development run. Aleks stored the task prompt and subsequent agent outputs in shared memory, allowing information generated at one stage to inform later agents in the development cycle (Jin et al., 2025). Once initiated, each run proceeded through its internal iterations without additional human input. The role and task prompts for both workflows are reproduced below. For privacy, a property-specific identifier in the Workflow 2 task prompt was replaced with bracketed text, and Files S1 and S2 contain the corresponding run logs with property-specific identifiers and local paths redacted.

For the model-development task in Workflow 1, Aleks completed separate 10 m, 3 m, and vine-scale runs using the vineyard dataset prepared for each prediction unit scale and 30 relevant publications listed in File S3. The role prompt supplied to all three Aleks runs was:

*You are an expert machine learning analyst. Your job is to summarize results, interpret performance, and provide suggestions.*

The retained task prompt was:

*The goal is to predict redvine disease in 2024 using data before Aug 2024. The resulting model will be used to predict future redvine disease. The presence column in the dataset is an average presence/absence of grape vine at grid points, no matter infected or not.*

The task prompt specified the prediction objective and temporal data boundary but left the full model-development approach for Aleks to determine. For the retained vine-scale run, redvine disease was operationalized as the presence of at least one 2024 red-leaf observation. The run was limited to 20 iterations. Researchers reviewed model designs between runs.

For the sampling plan design task in Workflow 2, Aleks received the retained vine-scale training script (File S4), the Workflow 1 run log (File S1), the 2025 vine-scale model outputs, the vine-scale vineyard dataset, and the 44 publications listed in File S3. The role prompt supplied to the sampling plan design Aleks run was:

*You are an expert in validation sampling for spatial machine learning models. You draw on statistical theory, sampling literature, and domain knowledge (spectroscopy) to design rigorous, unbiased validation strategies.*

The task prompt was:

*Validate the 2025 vine disease predictions. Select 50 vines from [the sampling vineyard] that will provide rigorous validation of model performance.*

The task prompt specified the sampling vineyard, sample size, and general goal of model validation but did not specify the intended comparison of virus detection rates between higher- and lower-scoring vines. The run was limited to five iterations and proceeded without intermediate human review.

## S2.4 – Workflow 1: Model development and evaluation

Aleks developed the final vine-scale model used to forecast 2025 red-leaf observations. We evaluated the 2025 forecasts and retrospectively compared five whole-row scouting strategies.

### S2.4.1 – Developing a vine-scale red-leaf forecasting model using Aleks

For the final vine-scale model, Aleks classified positions with at least one red-leaf record from September through December 2024 as observed positive. The dataset contained 9,026 observed-positive and 287,463 observed-negative positions. The model used 16 predictors (Table S4). The run log in File S1 records all 20 iterations. The total duration of the final run was 145 minutes. File S4 contains the model-development script from iteration 20 and the 2025 forecasting script. Domain scientists reviewed all 20 iterations of the final vine-scale run post hoc using the reasoning log, the final model-development script, and performance on independent 2025 scouting data. We received the completed 2025 scouting records after generating the 2025 forecasts and completing field sampling.

Aleks generated two predictors of change between July 2023 and July 2024, one for EVI and one for canopy area. It also calculated the mean red-leaf count among five neighboring vine positions separately for 2022 and 2023. Temporal EVI differences have precedent as predictors in remote-sensing classification (Halabuk et al., 2015; Kinnebrew et al., 2022). BallTree use identified the nearest coordinate records from all mapped vine positions using Euclidean distance applied to unprojected WGS84 latitude and longitude. For each vine, the first of the six closest coordinate records was treated as the focal vine and excluded, and the red-leaf counts of the remaining five were averaged, consistent with nearest-neighbor averaging used to construct spatial predictors (Liu et al., 2022).

Aleks assigned integer cultivar codes using scikit-learn's LabelEncoder. Other values stored as text were converted to numeric values, with values that could not be converted labeled missing. Aleks replaced missing numeric values with predictor medians calculated across all 296,489 positions before the training and evaluation split, although independent internal evaluation requires estimating those medians from the training subset alone (Perez-Lebel et al., 2022). To address poor positive-class recall, Aleks used the Synthetic Minority Over-sampling Technique (SMOTE) with default settings and a random state of 42 to generate synthetic positive-class records until the two outcome classes were equal in size (Chawla et al., 2002). SMOTE treated all predictors, including cultivar codes and coordinates, as numeric variables. For the final vine-scale model iteration, Aleks applied SMOTE before splitting the data into 80% training and 20% evaluation subsets, compromising their independence (Blagus & Lusa, 2015). The split was stratified by class and used the same random state, yielding 57,493 evaluation records per class. The evaluation subset included synthetic records and was 50% positive, compared with 3.0% in the original dataset. Aleks's internal evaluation therefore did not assess performance on independent observations with the original class distribution. To place predictors on a common scale before model fitting, Aleks applied StandardScaler to both subsets using means and standard deviations calculated from the training subset.

For the final iteration, Aleks used Auto-sklearn for automated model, preprocessing, and hyperparameter selection (Feurer et al., 2015), with a search time of 300 s, a memory limit of 4096 MB, and a random seed of 42. The script did not set the Auto-sklearn resampling method or model-selection metric. The exact Auto-sklearn version and complete software environment were not retained, limiting exact reconstruction of the model-development run. The final model score was Auto-sklearn's positive-class predict_proba output, ranging from 0 to 1, with higher values indicating greater model support for the observed-positive class. Scores of 0.50 or greater were classified positive. Aleks assessed its predictions for the evaluation subset using class-specific precision, recall, and F1 scores, a confusion matrix, and overall accuracy. Aleks reported the final model's internal performance for a single training/evaluation split (File S1), which can yield an unrepresentative performance estimate if results depend strongly on the chosen partition (Krstajic et al., 2014). Iteration 19 used class weighting and evaluated 57,493 observed-negative and 1,805 observed-positive records, whereas iteration 20 evaluated 57,493 records per class, including synthetic records. In proposing iteration 20, Aleks claimed that SMOTE handled class imbalance better than class weighting (File S1). The reported metrics do not provide a direct comparison because the evaluations used different class proportions and only the SMOTE evaluation included synthetic records. The fitted Auto-sklearn model and StandardScaler were saved for 2025 forecasting without refitting either object on the full dataset.

### S2.4.2 – Forecasting red-leaf observations for harvest 2025

To assess the operational value of Aleks's modeling deliverable, we forecast 2025 red-leaf observations using the fitted model resulting from Aleks's original preprocessing and SMOTE decisions (Section S2.4.1). We used the 2025 forecasting script in File S4 to prepare 2025 values for the 16 predictors used in the 2024 model by advancing the year-specific inputs by one year. Table S4 lists the complete substitutions.

For 2025 forecasting, the script fitted a new LabelEncoder to the unchanged cultivar labels. The original encoder had not been saved. A retrospective comparison confirmed that cultivar codes matched between model development and 2025 forecasting for all 296,489 vine positions. We replaced missing predictor values with the median of the same predictor across the full 2025 dataset. Across the management area, June and July 2025 EVI and canopy-area values were available for 53,148 (17.9%) and 125,926 (42.5%) of 296,489 positions, respectively. In the sampling vineyard, June 2025 values were unavailable and replaced with full-dataset medians. July 2024 and July 2025 values were complete for all 12,792 positions, so the annual EVI and canopy-area change predictors required no imputation.

After preparing the 2025 predictors, we applied the fitted StandardScaler and final model without refitting either object. Domain scientists received a CSV file containing one 2025 model score and corresponding predicted class for each of the 296,489 mapped vine positions.

### S2.4.3 – Evaluating 2025 red-leaf observation forecasts

We evaluated the 2025 red-leaf forecasts using the completed harvest-2025 visual scouting records. The "All in 2025" evaluation included all vine positions. The "New in 2025" evaluation included vine positions without a red-leaf record from 2021 through 2024. In both evaluations, a 2025 red-leaf record defined an observed positive, and the remaining eligible positions were observed negatives.

We calculated vine-scale model-performance metrics across the full management area and within the sampling vineyard described in Section S2.5. ROC AUC and average precision summarized how well model scores ranked observed-positive positions above observed-negative positions. ROC AUC was calculated from Mann–Whitney ranks, with tied scores assigned their average rank. Average precision was calculated as the mean precision at observed-positive ranks after ordering model scores from highest to lowest. Average-precision enrichment was calculated by dividing average precision by observed-positive prevalence. Median model scores and 25th–75th percentile ranges were calculated separately for observed-positive and observed-negative positions.

To assess classification at the 0.50 threshold, positions with scores of 0.50 or greater were classified as predicted positive and positions with lower scores as predicted negative. Precision was the proportion of predicted-positive positions that were observed positive, and recall was the proportion of observed-positive positions classified as predicted positive. Precision enrichment was calculated by dividing precision at the 0.50 threshold by observed-positive prevalence. Results for the full management area are reported in Table 1, and expanded results for the full management area and sampling vineyard are reported in Table S1.

To assess model score ranking within blocks, we calculated block-adjusted ROC AUC for the full management area following Janes et al. (2009). For each observed-positive vine, we determined the proportion of observed-negative vines in the same ranch and block with a lower model score, with half credit for ties. We averaged these proportions across observed-positive vines in blocks containing both observed-positive and observed-negative vines. Vine-scale model-performance metrics were calculated in R 4.5.2 using custom base R functions.

### S2.4.4 – Retrospective comparison of whole-row scouting strategies

We retrospectively compared recall across survey effort for the five whole-row scouting strategies described in main-text Section 2.4.3. Figure 2 reports recall across survey effort, and Figure S1 reports recall differences and recall exceedance for the dataset-wide and within-block random comparisons. Table S2 reports survey effort, recall, and dataset-wide comparisons at 10%, 20%, and 45% survey effort; Table S3 reports the corresponding within-block random comparisons.

We defined each vineyard row by its ranch, block, and row identifiers. Survey effort was the percentage of all vine positions contained within surveyed rows, and recall was the percentage of vine positions with newly recorded 2025 red-leaf observations contained within those rows. At each target effort, we simulated surveying whole rows until the total number of surveyed vine positions reached the target. Realized survey effort could therefore exceed target effort. We evaluated survey effort from 0.25% to 45% in 0.25-percentage-point increments. Systematic and adaptive scouting skipped rows, so some routes ended after surveying approximately half of all vine positions. We therefore stopped at 45%, the highest survey effort reached consistently by all five strategies.

We constructed random, systematic, and adaptive scouting routes by varying row order, survey direction, and starting position. A route was the ordered sequence of rows produced by one set of these choices. Random scouting ordered all rows across the full management area randomly without regard to block. Systematic scouting randomized block order and survey direction, began with either the first or second row in each block, and then surveyed every other row. Adaptive scouting randomized block order and survey direction, began with the first, second, or third row in each block, and initially surveyed every third row. Within each replicate, systematic and adaptive scouting shared the randomized block order and survey direction. We selected the starting position of each route independently. For systematic and adaptive scouting, survey direction and starting position were the same across blocks within each route.

We constructed the two model-based scouting strategies by assigning each row the mean model score of its highest-scoring 10% of vine positions. The number of vine positions included in this calculation was rounded up, with a minimum of one per row. Model-prioritized scouting selected rows in descending order of their row scores. Model-adaptive scouting used the same ranking together with the adaptive expansion rule described below.

Adaptive and model-adaptive scouting used the same expansion rule. Any 2025 red-leaf observation in an initially selected or model-ranked row triggered the addition of up to two previously unsurveyed neighboring rows on each side within the same block. The simulation checked a row for a 2025 red-leaf observation only when that row was surveyed. Recall nevertheless counted only vine positions with newly recorded 2025 red-leaf observations. Observations in added rows did not trigger further expansion. If a target effort was reached before every neighboring row added by a trigger had been surveyed, the remaining neighboring rows were not surveyed.

To characterize variation among routes, we generated 5,000 routes each for random, systematic, and adaptive scouting across the full management area. Random seeds were 20260718 for random scouting and 20250711 for systematic and adaptive scouting. Model-prioritized and model-adaptive scouting each produced one route because their rankings and expansion rules contained no random choices. For the dataset-wide random comparisons, each of the 5,000 systematic routes and each of the 5,000 adaptive routes was paired with one independently generated random route at the same target effort. Each model-based route was compared with all 5,000 random routes. We compared the two model-based routes directly. For systematic and adaptive scouting, we separately calculated the percentage of routes with greater recall than the model-adaptive route. Recall exceedance was the percentage of comparisons in which a strategy achieved strictly greater recall than its comparator.

To assess row selection while holding block allocation constant, we used within-block random comparisons. For each comparison, we randomly selected rows without replacement, matching the number of rows from each block in the corresponding strategy route. We generated 5,000 comparisons for each strategy and target effort using random seed 20250714. Rows contained different numbers of vine positions, so preserving row counts did not always preserve realized survey effort exactly.

The simulations measured vine-position coverage and did not account for travel distance, elapsed time, or labor. The reported 2.5th–97.5th percentile ranges described variation among simulated routes rather than uncertainty across vineyards or years. Monte Carlo standard errors quantified simulation precision and were not calculated for comparisons involving only deterministic routes. Analyses used R 4.5.2.

## S2.5 – Workflow 2: Sampling plan design and evaluation

Domain scientists and Aleks independently developed sampling plans for the sampling vineyard. We implemented both plans, submitted samples for molecular testing, and compared the selected positions and sampling outcomes. We assessed agreement between petiole and cane test results and evaluated associations between December cane results and model scores.

### S2.5.1 – Sampling plan development

We restricted sampling for virus testing to one Cabernet Sauvignon vineyard to limit variation in cultivar and management and simplify navigation to individual vines. We selected the vineyard for its uniform layout, continuous recordkeeping, sufficient candidates in both model-predicted classes, and a model score gradient extending from a historical red-leaf hotspot to a lower-scoring area (Figure 3). Both sampling plans drew from the sampling vineyard's 12,792 vine positions, with 8,364 in one block and 4,428 in the other. We independently developed a 100-position domain-scientist plan, while Aleks was instructed to develop a 50-position plan over a maximum of five design iterations.

Aleks received two tabular inputs covering all 296,489 mapped vine positions, in addition to the prompts and literature described in Section S2.3. The first was the 2025 prediction output generated as described in Supplementary Methods Section S2.4.2. It contained vine identifiers and coordinates, ranch and block identifiers, the predicted class, the model score, and the 2024 red-leaf count for each position. The second was the 75-variable vineyard dataset used to prepare the model inputs. Aleks filtered both tabular inputs to the 12,792 positions in the sampling vineyard before developing the plan.

Aleks proposed four sampling designs across five iterations during a 15-minute run (File S2). It generated Python code for each design, with up to three code-generation attempts, and assessed selections using summaries of sample size, uniqueness, predicted classes, model scores, spatial coverage, and pairwise distances. The final design targeted 10 positions from each of five model score quintiles and a minimum spacing of 15 m. Aleks proposed randomly ordering 15 × 15 m grid cells, selecting no more than one eligible position per cell, and excluding the selected cell and a 15 m buffer around each selected position from further selection. This emphasis on spatial coverage has precedent in vineyard sampling (Robertson et al., 2022). The design also specified checks of block representation and variation in annual EVI and canopy-area changes.

The final positions differed from the proposed quintile allocation (Table 2), and Aleks reported a minimum spacing of 5.5 m (File S2). Domain scientists received the reasoning log and final selected positions, which supported review of the stated design and final plan. The executable selection script and intermediate plan files were not retained, so we could not reproduce the exact selection process.

For the domain-scientist plan, candidate positions had no 2024 red-leaf record. Positions with records from 2021-2023 remained eligible. We divided candidates into four subclasses according to whether their model score was below or at least 0.50 and whether they were within 3 m of a 2024 red-leaf observation. We selected 25 positions from each subclass, yielding 100 positions across 10 of the sampling vineyard's 78 rows. We semi-randomly selected rows spaced to span the sampling vineyard and represent the remaining positions still missing for each subclass. Within each row, we aimed to include as many subclasses as possible and randomly selected positions within those subclasses, totaling 10 positions per row. The minimum spacing among planned positions was 1.8 m (neighboring vines).

### S2.5.2 – Field implementation and virus testing

We located planned positions using row and vine counts, with grid alignment confirmation supported by a VineView PinPoint real-time kinematic device. On 16–17 September 2025, we collected petiole tissue from 97 domain-scientist-plan vines. For each vine, we collected the two most basal leaf petioles from each of the six canes closest to the trunk, yielding six petioles from each cordon. Between the 9th and 10th of December 2025, we collected cane tissue from the same 97 domain-scientist-plan vines and from 41 of the 50 Aleks-plan positions. We pooled six canes from each vine into one cane sample. The same researcher implemented the domain-scientist petiole-sampling plan and the Aleks cane-sampling plan.

During September petiole sampling, we selected replacement positions when a selected position was empty or contained a vine younger than three years. We randomly selected replacements from the same sampling subclass in selected rows that had not yet been sampled. We excluded young vines to reduce the chance that historical records at a sampled position described a previous vine. The model inputs did not track vine removal or replanting. Three domain-scientist targets remained unfilled after in-field replacements. Nine Aleks-plan positions were empty or contained vines younger than three years during December cane sampling. The Aleks plan did not provide replacement instructions, so those positions remained unsampled.

We collected tissue with pruners, changed gloves between vines, and placed material from each vine in a separate resealable plastic bag. We transported the samples on ice to AgriAnalysis (Davis, California, USA) for testing of grapevine red blotch virus (GRBV) and grapevine leafroll-associated virus 3 (GLRaV-3). The laboratory used reverse transcription polymerase chain reaction (RT-PCR) for GRBV and an enzyme-linked immunosorbent assay (ELISA) with in-house reagents for GLRaV-3. December testing provided results for 97 scientist-plan positions and 41 Aleks-plan positions, including one shared vine, for 137 unique vines. We included the shared vine in each plan's summary and counted it once in analyses combining the plans.

### S2.5.3 – Sampling plan and model score evaluation

We compared the plans' model score coverage, spatial distribution of selected positions, and field sampling outcomes (Table 2). For each plan, we calculated the percentage of selected positions with scores of at least 0.50 and the percentage in each quintile of the sampling vineyard's model score distribution. We described spatial distribution using the percentage of selected positions within 3 m of a 2024 red-leaf observation, the minimum spacing between selected positions, and the number of rows represented. We summarized field sampling outcomes as the percentage of planned positions tested, the median number of tested vines per sampled row, and counts of GRBV-positive and GRBV-negative cane samples.

We compared September petiole and December cane results for the 97 domain-scientist-plan vines. For each virus, we calculated the percentage of vines with matching test classifications. We used December cane results for model score analyses because cane tissue was collected under both plans. We combined results across plans, counting the shared vine once. We calculated recall as the percentage of GRBV-positive vines with model scores of at least 0.50. Within each model score group (at least 0.50 or below 0.50), we calculated the percentage of vines testing positive for GRBV. We also summarized median model scores for GRBV-positive and GRBV-negative vines. We calculated ROC AUC as pairwise concordance between the five GRBV-positive and 132 GRBV-negative vines, assigning half credit to tied scores. Calculations were completed using R 4.5.2. Results are reported in Table 3.

# Supplemental Figures and Tables

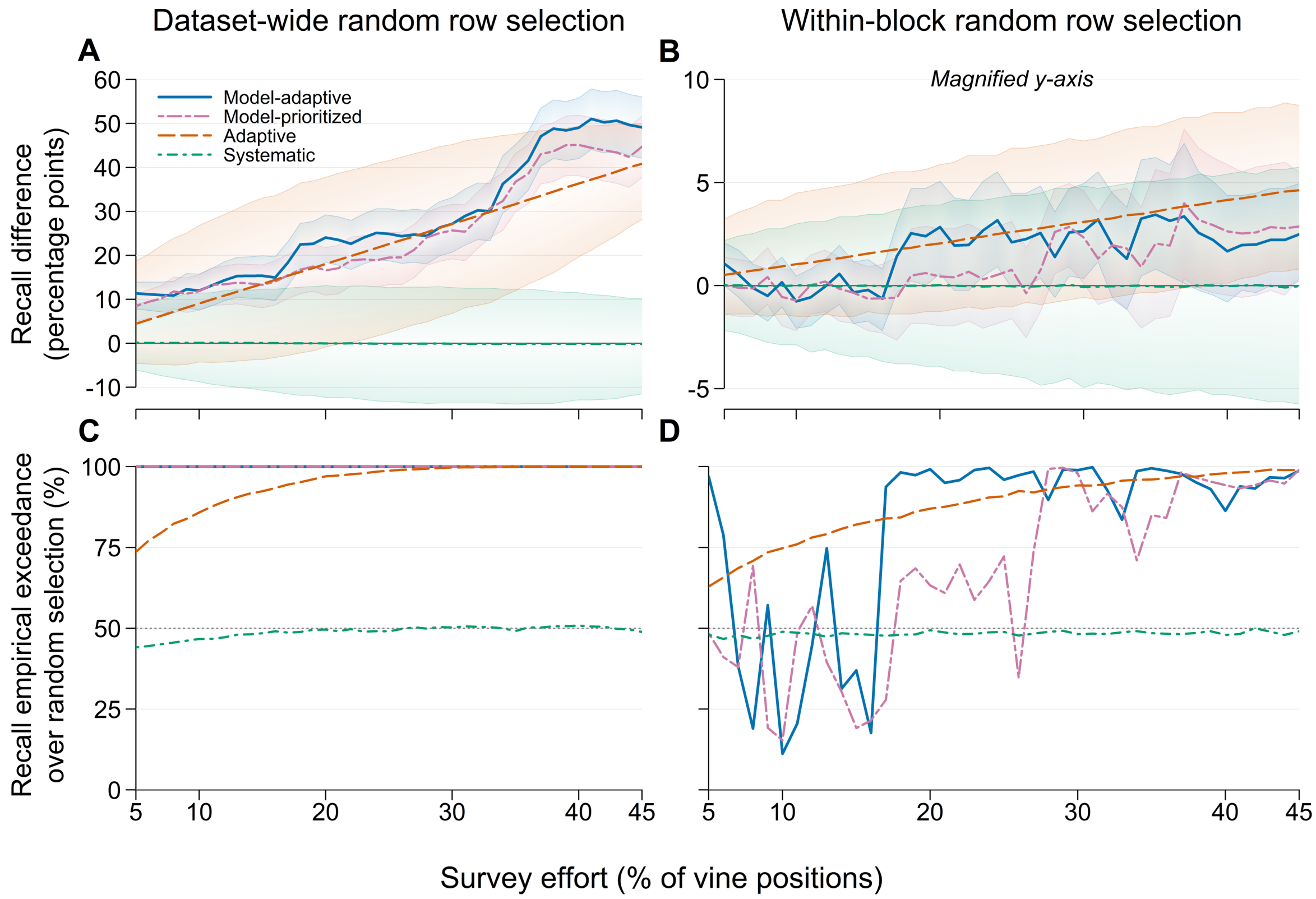

**Figure S1. Recall comparisons of four whole-row scouting strategies with dataset-wide and within-block-matched random row selection.** Recall was the percentage of vine positions with newly recorded 2025 red-leaf observations within surveyed rows. All panels show 5%–45% survey effort in 1-percentage-point increments. A and C compare each strategy with dataset-wide random row selection across the 140-ha management area; B and D compare each route with a within-block random selection containing the same number of rows per block. Matching row counts did not always preserve realized survey effort exactly. Upper panels show mean strategy-minus-random recall differences in percentage points; ribbons span the 2.5th–97.5th percentiles across 5,000 comparisons. Lower panels show recall exceedance, the percentage of comparisons in which strategy recall was strictly greater than random-selection recall. Tables S2 and S3 report Monte Carlo standard errors as well as corresponding dataset-wide and within-block results at 10%, 20%, and 45% survey effort.

**Table S1. 2025 vine-scale model performance for the full dataset and sampling vineyard.** "All in 2025" includes all vine positions. "New in 2025" includes positions without red-leaf records during 2021-2024. Observed + and observed − indicate positions with and without a 2025 red-leaf record, respectively. Observed + percentages use all evaluated positions in each column. Median model scores are followed by 25th-75th percentile ranges. Scores ≥0.50 define predicted + positions, and lower scores define predicted − positions. Precision is the percentage of predicted + positions that were observed +. Recall is the percentage of observed + positions that were predicted +. Enrichment is average precision or precision at 0.50 divided by observed + prevalence. Block-adjusted ROC AUC evaluates score ranking within blocks containing both observed classes (Janes et al., 2009). For the full dataset, the 44 contributing blocks in "All in 2025" and 42 in "New in 2025" contained all observed + positions and 38.94% and 39.59% of evaluated positions, respectively. Supplementary Methods Section S2.4.3 describes metric calculations.

| | **Full dataset – All in 2025** | **Full dataset – New in 2025** | **Sampling vineyard - All in 2025** | **Sampling vineyard - New in 2025** |
|---|---|---|---|---|
| Evaluated positions | 296,489 | 251,749 | 12,792 | 10,910 |
| Observed + positions (%) | 1,163 (0.39) | 869 (0.35) | 196 (1.53) | 154 (1.41) |
| Average precision (%) | 0.70 | 0.65 | 2.39 | 2.27 |
| Average precision enrichment (× prevalence) | 1.79x | 1.89x | 1.56x | 1.61x |
| ROC AUC | 0.667 | 0.659 | 0.615 | 0.613 |
| Median score, observed + | 0.263 (0.200-0.318) | 0.260 (0.197-0.306) | 0.363 (0.292-0.467) | 0.349 (0.286-0.456) |
| Median score, observed - | 0.197 (0.180-0.275) | 0.187 (0.179-0.271) | 0.310 (0.257-0.410) | 0.308 (0.256-0.399) |
| Precision at 0.50, (%) | 45/4,140 (1.09) | 32/2,223 (1.44) | 34/1,345 (2.53) | 25/1,020 (2.45) |
| Precision enrichment at 0.50 (× prevalence) | 2.77x | 4.17x | 1.65x | 1.74x |
| Recall at 0.50, (%) | 45/1,163 (3.87) | 32/869 (3.68) | 34/196 (17.35) | 25/154 (16.23) |
| Block-adjusted ROC AUC | 0.57 | 0.52 | | |

**Table S2. Survey effort, recall, and route comparisons for five whole-row scouting strategies in the 140-ha management area.** Random scouting selected rows across the whole management area. Recall was the percentage of vine positions with newly recorded 2025 red-leaf observations within surveyed rows. At 10%, 20%, and 45% survey effort, mean realized survey effort exceeded target effort by less than 0.04 percentage points for every strategy. Recall for random, systematic, and adaptive scouting is reported as means across 5,000 simulated routes, with 2.5th–97.5th percentiles in parentheses. Each model-based strategy produced one route. Recall exceedance is the percentage of comparisons in which strategy recall was strictly greater than that of the random or model-adaptive comparator. Monte Carlo standard errors (MCSE) quantify simulation precision.

| Target effort (%) | **Scouting method** | **Recall, %** | **Recall exceedance vs random, % (MCSE)** | **Recall exceedance vs model-adaptive, % (MCSE)** |
|---|---|---|---|---|
| 10 | Model-prioritized | 21.86 | 100.0 (0.00) | 0 |
| | Model-adaptive | 21.98 | 100.0 (0.00) | |
| | Adaptive | 19.07 (6.33-36.36) | 85.8 (0.49) | 33.7 (0.67) |
| | Systematic | 10.12 (2.53-20.60) | 46.7 (0.71) | 1.2 (0.16) |
| | Random | 10.01 (6.21-14.38) | | |
| 20 | Model-prioritized | 36.59 | 100.0 (0.00) | 0 |
| | Model-adaptive | 44.07 | 100.0 (0.00) | |
| | Adaptive | 38.08 (19.91-57.08) | 97.0 (0.24) | 26.8 (0.63) |
| | Systematic | 20.01 (9.09-31.65) | 49.6 (0.71) | 0.0 (0.00) |
| | Random | 20.02 (14.84-25.66) | | |
| 45 | Model-prioritized | 89.76 | 100.0 (0.00) | 0 |
| | Model-adaptive | 94.13 | 100.0 (0.00) | |
| | Adaptive | 85.85 (75.26-90.91) | 100.0 (0.00) | 0.0 (0.00) |
| | Systematic | 44.86 (35.79-52.47) | 48.8 (0.71) | 0.0 (0.00) |
| | Random | 45.02 (38.09-52.01) | | |

**Table S3. Recall of four whole-row scouting strategies relative to within-block random row selection at 10%, 20%, and 45% survey effort in the 140-ha management area.** Recall was the percentage of vine positions with newly recorded 2025 red-leaf observations within surveyed rows. Mean survey effort in the random selections ranged from 0.383 percentage points below to 0.217 above the corresponding strategy. Strategy recall is reported as means across 5,000 systematic or adaptive routes and as single-route values for deterministic model-based strategies. Within-block random recall is reported as the mean across 5,000 matched selections, with 2.5th–97.5th percentiles in parentheses. Each selection matched the corresponding scouting route's row counts in each block. Recall exceedance is the percentage of matched comparisons in which strategy recall was strictly greater than within-block random recall. Monte Carlo standard errors (MCSE) quantify simulation precision.

| Target effort (%) | **Scouting strategy** | **Strategy recall (%)** | **Within-block random recall (%)** | **Recall exceedance, % (MCSE)** |
|---|---|---|---|---|
| 10 | Model-adaptive | 21.98 | 22.75 (21.52-24.05) | 10.8 (0.44) |
| | Model-prioritized | 21.86 | 22.57 (21.06-24.05) | 16.0 (0.52) |
| | Adaptive | 19.07 | 18.03 (6.10-34.64) | 75.8 (0.61) |
| | Systematic | 10.12 | 10.12 (2.53-20.71) | 48.3 (0.71) |
| 20 | Model-adaptive | 44.07 | 41.16 (38.90-43.38) | 99.3 (0.12) |
| | Model-prioritized | 36.59 | 36.17 (33.95-38.55) | 62.9 (0.68) |
| | Adaptive | 38.08 | 36.01 (18.64-54.20) | 86.5 (0.48) |
| | Systematic | 20.01 | 20.01 (9.09-32.11) | 49.9 (0.71) |
| 45 | Model-adaptive | 94.13 | 91.65 (89.18-93.90) | 98.8 (0.15) |
| | Model-prioritized | 89.76 | 86.85 (84.23-89.30) | 99.0 (0.14) |
| | Adaptive | 85.85 | 81.20 (70.08-87.11) | 99.4 (0.11) |
| | Systematic | 44.86 | 44.88 (35.33-52.70) | 48.7 (0.71) |

**Table S4. Predictors and inputs used to develop the final vine-scale model and forecast 2025 red-leaf symptoms.** Model-development inputs covered 2021-2024 and were aligned with 2024 red-leaf outcomes. For 2025 forecasting, year-specific inputs were advanced by one year, whereas longitude, latitude, cultivar, and vine spacing were unchanged.

| Predictor | 2024 model input | 2025 forecast input |
|---|---|---|
| Red-leaf record count 1 | 2021 | 2022 |
| Red-leaf record count 2 | 2022 | 2023 |
| Red-leaf record count 3 | 2023 | 2024 |
| Girdled-vine count | 2023 | 2024 |
| Early-summer EVI | Jun 4, 2024 | Jun 5, 2025 |
| Midsummer EVI | Jul 10, 2024 | Jul 8, 2025 |
| Early-summer canopy area | Jun 4, 2024 | Jun 5, 2025 |
| Midsummer canopy area | Jul 10, 2024 | Jul 8, 2025 |
| Longitude | Vine longitude | Unchanged |
| Latitude | Vine latitude | Unchanged |
| Cultivar | Cultivar assigned to each block | Unchanged |
| Vine spacing | Vine spacing | Unchanged |
| Annual EVI change | Jul 10, 2024 minus Jul 30, 2023 | Jul 8, 2025 minus Jul 10, 2024 |
| Annual canopy-area change | Jul 10, 2024 minus Jul 30, 2023 | Jul 8, 2025 minus Jul 10, 2024 |
| Earlier neighborhood mean | Mean 2022 red-leaf count among five nearest vine positions | Mean 2023 red-leaf count among five nearest vine positions |
| Later neighborhood mean | Mean 2023 red-leaf count among five nearest vine positions | Mean 2024 red-leaf count among five nearest vine positions |

## Supplemental References